\documentclass[twocolumn, nature, superscriptaddress]{revtex4-2}

\usepackage{amsmath}
\usepackage{amsbsy}
\usepackage{amssymb}
\usepackage{amstext}
\usepackage{mathtools}
\usepackage{graphicx}
\usepackage[unicode=true, pdfusetitle, bookmarks=false, breaklinks=false, pdfborder={0 0 1}, backref=false, colorlinks=false]{hyperref}
\hypersetup{colorlinks=true,citecolor=NavyBlue,filecolor=black,linkcolor=black,urlcolor=black}
\usepackage{txfonts,color}
\usepackage[dvipsnames]{xcolor}
\usepackage[normalem]{ulem}
\usepackage[latin9]{inputenc}

\newcommand{\benremove}[1]{{\color[rgb]{0,0,0.9}\sout{#1}}}
 
\renewcommand{\benremove}[1]{} 

\begin{document}
\title{Detection of Molecular Transitions with Nitrogen-Vacancy Centers and Electron-Spin Labels}
\author{C. Munuera-Javaloy${}^*$}
\affiliation{Department of Physical Chemistry, University of the Basque Country UPV/EHU, Apartado 644, 48080 Bilbao, Spain}
\affiliation{EHU Quantum Center, University of the Basque Country UPV/EHU, Leioa, Spain.}
\author{R. Puebla}
\affiliation{Instituto de F\'{i}sica Fundamental, IFF-CSIC, Calle Serrano 113b, 28006 Madrid, Spain}
\affiliation{Centre for Theoretical Atomic, Molecular, and Optical Physics, School of Mathematics and Physics, Queen's University, Belfast BT7 1NN, United Kingdom}
\author{B. D'Anjou}
\affiliation{Institut f\"ur Theoretische Physik and IQST, Universit\"at Ulm, Albert-Einstein-Allee 11, D-89069 Ulm, Germany}
\author{M. B. Plenio}
\affiliation{Institut f\"ur Theoretische Physik and IQST, Universit\"at Ulm, Albert-Einstein-Allee 11, D-89069 Ulm, Germany}
\author{J. Casanova}
\affiliation{Department of Physical Chemistry, University of the Basque Country UPV/EHU, Apartado 644, 48080 Bilbao, Spain}
\affiliation{EHU Quantum Center, University of the Basque Country UPV/EHU, Leioa, Spain.}
\affiliation{IKERBASQUE, Basque Foundation for Science, Plaza Euskadi 5, 48009 Bilbao, Spain}

\begin{abstract}
We present a protocol that detects molecular conformational changes with two nitroxide electron-spin labels and a nitrogen-vacancy (NV) center in diamond. More specifically, we demonstrate that the NV can detect energy shifts induced by the coupling between electron-spin labels. The protocol relies on the judicious application of microwave and radiofrequency pulses in a range of parameters that ensures stable nitroxide resonances. Furthermore, we demonstrate that our scheme is optimized by using nitroxides with distinct nitrogen isotopes. We develop a simple theoretical model that we combine with Bayesian inference techniques to demonstrate that our method enables the detection of conformational changes in ambient conditions including strong NV dephasing rates as a consequence of the diamond surface proximity and nitroxide thermalization mechanisms. Finally, we counter-intuitively show that with our method the small residual effect of random molecular tumbling becomes a resource that can be exploited to extract inter-label distances.
\end{abstract}

\maketitle
\section*{INTRODUCTION}
A precise knowledge of molecular structure and dynamics would shed new light on the mechanisms by which, e.g., proteins interact with their environment~\cite{Stetz19}. This finds important applications in distinct fields such as biochemistry and biology, since macromolecular conformational changes underlie numerous biological processes. In this regard, different techniques have been proposed to study molecular structure and dynamics~\cite{Rhodes06, Duddeck98, Alonso14}. In particular, methods based on nuclear magnetic resonance (NMR)~\cite{Abragam61, Levitt08} and on electron-spin resonance (ESR)~\cite{Prisner01,Schweiger01} stand out as they are central to investigate a wide variety of molecules, with ESR being more sensitive due to the much larger gyromagnetic ratio of electrons~\cite{Borbat01}. One commonly used ESR method is the Double Electron-Electron Resonance (DEER) in which synchronized and/or delayed pulses are delivered to, e.g., estimate average distances among free radicals in molecular ensembles~\cite{Jeschke12}. However, traditional ESR based on inductive detection with macroscopic coils requires macroscopic samples with a high chemical purity to produce a detectable response. This rules out ESR investigations of static and dynamic properties of individual molecules.

Detectors based on NV centers in diamond~\cite{Doherty13, Wu16} are promising candidates to overcome the limitations of inductive detection. Indeed, such detectors enable ESR and NMR detection of molecular samples down to the micron- and nanoscale and even down to the level of single molecules. The potential of NV centers in diamond originates from
a number of appealing properties and early demonstrations. They can be initialized and read out using
laser light in the visible spectrum, thereby removing thermal fluctuations from the measurement process~\cite{Manson06, Steiner10, Waldherr11}. Moreover, NV hyperfine states can be manipulated with microwave radiation in such a way that the NV couples coherently to nearby nuclear and electron spins while at the same time being decoupled from environmental fluctuations~\cite{cai2013diamond,cai2012robust,london2013detecting,Wang19,Munuera21}.
These concepts have enabled shallowly implanted NV centers to detect electron spins
\cite{grotz2011sensing} and even nuclear spins ~\cite{Mueller14,Sushkov14} above the diamond surface with single-spin sensitivity. Detection protocols have also been developed allowing for nanoscale NMR~\cite{schmitt2017submillihertz,Glenn18,Bucher20,Arunkumar21} and ESR detection~\cite{chu2020proposal,meinel2021heterodyne,staudenmaier2021phase} with spectral resolution approaching that of macroscopic ESR and NMR based on inductive detection. Fluctuating electron spins of biological samples~\cite{ermakova2013detection,ziem2013highly,rendler2017optical,barton2020nanoscale} and small groups of statistically polarized nuclear spins in a target protein \cite{Lovchinsky16} have been detected. Finally and particularly relevant to the present work, an NV center has also been used to detect a single {\it nitroxide electron-spin label}~\cite{Shi15} attached to a protein. The use of nitroxide spin labels is particularly attractive because it enables the study of non-paramagnetic molecules~\cite{Haugland16}. In particular, Ref.~\cite{Schlipf17} reported the detection of an effective inter-label coupling in molecular ensembles using advanced multi-frequency spectroscopy techniques. Attaching several nitroxides to a single target protein could enable the study of its internal dynamics at room temperature with an NV center. However, this approach is challenging because spin labels have resonance frequencies and interaction strengths that strongly depend on their orientation with respect to the external magnetic field~\cite{Marsh15}. Therefore, the identification of these parameters is complicated by the unavoidable molecular motion and environmental noise affecting the NV and the electron-spin labels.

In this work, we show how even a simple pulse sequence applied to a single pair of electron-spin labels and a shallow NV [see Fig.~\ref{fig:system}(a)] can detect single-molecule conformational transitions at room temperature. Our carefully engineered DEER sequence asymmetrically delivers microwave (radiofrequency) pulses to electron spins of the NV center (nitroxide spin labels) to resolve small energy shifts in the NV spectrum owing to the coupling between labels. Crucially, we find an energy-transition branch of the labels that presents an excellent robustness against unavoidable tumbling motion in ambient conditions. In addition, we show that this branch is especially robust when each nitroxide label contains a distinct nitrogen isotope. Finally, we develop a simple but accurate model which enables us to estimate not only the coupling but also the distance between labels using Bayesian inference. In particular, we counter-intuitively show that the small residual effect of tumbling, instead of being detrimental, enables the extraction of the distance between labels.

\section*{RESULTS}
\subsection*{The System}
\begin{figure}[t]
\hspace{0.0 cm}\includegraphics[width=1.0\columnwidth]{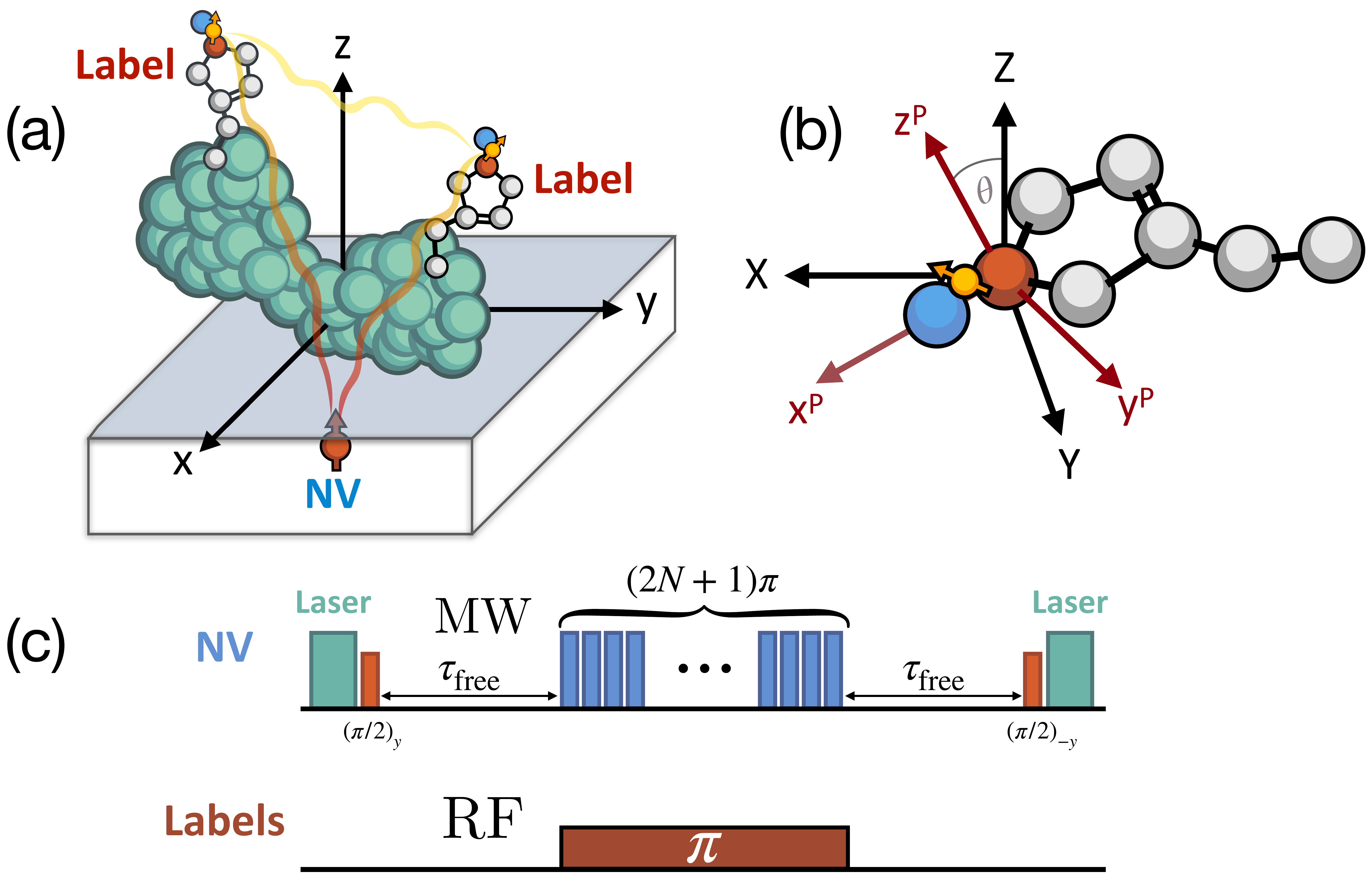}
\caption{(a) A protein with two attached labels is near a diamond surface that contains a shallow NV. (b) Nitroxide molecule including the electron spin (yellow). The principal axes are shown in red, while the laboratory frame is shown in black. The azimuth $\theta$ (the angle between the z principal axis and the z laboratory axis) determines the energy-transition branches of the label. (c) MW and RF driving scheme. This includes initialization and readout with laser and microwave pulses, two free evolution stages of duration $\tau_{\rm free}$, $2N+1$ MW $\pi$-pulses on the NV, and a simultaneous RF $\pi$-pulse on the labels. \label{fig:system}}
\end{figure}

The Hamiltonian of the NV, the two nitroxide electron-spin labels, and the microwave (MW) and radiofrequency (RF) driving fields reads
\begin{align}
\begin{aligned}
H = &\frac{1}{2}\left(\mathbb{I} + \sigma^z \right)\left(\vec{a}_1 \cdot \vec{J}_1+\vec{a}_2 \cdot \vec{J}_2\right) \\
&+ H_{n_1}+H_{n_2} + g_{12}\left[J_1^z J_2^z-\frac{1}{4}\left(J_1^+ J_2^- + J_1^- J_2^+ \right)\right] \\
&+\frac{\Omega_{\rm MW}}{2} \sigma^x+2 \Omega_{\rm RF} \left(J_1^x+J_2^x\right) \cos(\omega_{\rm RF} t). \label{eq:main}
\end{aligned}
\end{align}
Here, $\sigma^z$ is the Pauli $z$ operator for the NV two-level system and $\vec{J}_i$ are spin-$1/2$ operators for each label ($i=1,2$). The Hamiltonian of the $i$th nitroxide label is $H_{n_i}$ and the coupling of each label to the NV is mediated by the vectors $\vec{a}_i$. Moreover, $J_i^\pm = J_i^x \pm i J_i^y$ are electron-spin ladder operators and $g_{12}$ is the coupling constant between labels. The last line in Eq.~\eqref{eq:main} describes a MW field resonant with the NV, leading to the term $\frac{\Omega_{\rm MW}}{2} \sigma^x$ in a suitable interaction picture. In addition, an RF field of frequency $\omega_{\rm RF}$ and Rabi frequency $\Omega_{\rm RF}$ excites the electronic resonances of the nitroxide labels. Equation~\eqref{eq:main} is derived in the Supplemental Material~\cite{SM} and a sketch of the system is presented in Fig.~\ref{fig:system}(a).

The term $H_{n_i}$ models each nitroxide label as~\cite{Marsh15, Marsh19}
\begin{align}
\begin{aligned}
H_{n_i} = \mu_B B^z \hat{z}\cdot \mathbb{G}_i \cdot \vec{J}_i+ \gamma_N B^z I_i^z+\vec{I}_i \cdot \mathbb{Q}_i \cdot \vec{I}_i+\vec{J}_i \cdot \mathbb{A}_i \cdot \vec{I}_i, \label{eq:nitro}
\end{aligned}
\end{align}
where $\mu_B$ is the Bohr magneton, $B^z$ is a magnetic field applied along the $z$ axis, $\gamma_{\rm N}$ is the nuclear gyromagnetic ratio equal to $2\pi \times 3.077$\ kHz/mT for $^{14}$N and to $2\pi \times -4.316$\ kHz/mT for $^{15}$N, $\vec{I}_i$ is the nuclear spin operator for the $i$th nitroxide label, and $\mathbb{G}_i$, $\mathbb{Q}_i$, and $\mathbb{A}_i$ are respectively the Land\'e, quadrupolar, and electron-nucleus interaction tensors~\cite{Marsh19}. We note that $\vec{I}_i$ is a spin-1 (spin-$1/2$) operator when describing a $^{14}$N ($^{15}$N) nuclear spin. The components of the $\mathbb{G}_i$, $\mathbb{Q}_i$, and $\mathbb{A}_i$ tensors in a general reference frame depend on the frame's relative orientation with respect to the principal axes of the nitroxide [see Fig.~\ref{fig:system}(b)]. In the principal frame, $\mathbb{G}_i$, $\mathbb{Q}_i$, and $\mathbb{A}_i$ are diagonal~\cite{Marsh19}. In particular, the Land\'e tensor in the principal frame is $\mathbb{G}_i^{(P)} = {\rm diag}\left(G^x,G^y,G^z\right)$, with $G^x\approx G^y\approx 2.007 \equiv G^\perp$ and $G^z \approx 2.002 \equiv G^\parallel$. The quadrupolar tensor for $^{14}$N is $\mathbb{Q}_i^{(P)} = {\rm diag}\left(Q^x,Q^y,Q^z\right)$, with $Q^x \approx 2\pi \times 1.26$ MHz, $Q^y \approx 2\pi \times 0.53$ MHz, and $Q^z \approx 2\pi \times -1.79$ MHz. The quadrupolar tensor vanishes for $^{15}$N. Finally, the tensor that mediates electron-nucleus interactions in each nitroxide is $\mathbb{A}_i^{(P)} = {\rm diag}\left(A^x,A^y,A^z\right)$, where $A^x\approx A^y\approx 2\pi \times 14.7 \ {\rm MHz} \equiv A^\perp$ and $A^z \approx 2\pi \times 101.4 \ {\rm MHz} \equiv A^\parallel$ for $^{14}$N, while $A^x\approx A^y\approx 2\pi \times 27 \ {\rm MHz} \equiv A^\perp$ and $A^z \approx 2\pi \times 141 \ {\rm MHz} \equiv A^\parallel$ for $^{15}$N~\cite{Marsh19}.

We obtain approximate nitroxide transition frequencies via a perturbative treatment of the nuclear degrees of freedom in Eq.~(\ref{eq:nitro})~\cite{SM}. For a nitroxide hosting
$^{14}$N, we find
\begin{align}
\begin{aligned}
H_{n_i} \approx \left[ E_1^i |\widetilde{1}\rangle\langle \widetilde{1}|_i + E_0^i |\widetilde{0}\rangle\langle \widetilde{0}|_i + E_{-1}^i |-\widetilde 1\rangle\langle -\widetilde{1}|_i \right] J_i^z. \label{eq:trace1}
\end{aligned}
\end{align}
The three {\it energy-transition branches} exhibit transition energies given by
$E_{1,-1}^i= \mu_B B^z G(\theta_i) \pm \frac{1}{\sqrt{2}} \sqrt{\left[(A^{\parallel})^2- (A^{\perp})^2\right] \cos (2 \theta_i )+(A^{\perp})^2+(A^{\parallel})^2}$
 and $E_0^i= \mu_B B^z G(\theta_i) + \frac{1}{2 \mu_B B^z G(\theta_i)} \frac{2 (A^{\perp} A^{\parallel})^2 +\left[(A^{\perp})^4-(A^{\perp} A^{\parallel})^2\right] \sin^2(\theta_i)}{(A^{\perp})^2 \sin^2(\theta_i)+(A^{\parallel})^2 \cos^2(\theta_i)}$.
Here, $\theta_i$ is the azimuth angle between the applied magnetic field and the principal $z$ axis of the nitroxide [see Fig. \ref{fig:system}(b)], $G(\theta_i) =\frac{1}{2} [(G^\parallel - G^\perp) \cos (2 \theta_i)+ G^\perp+ G^\parallel]$, and $| \widetilde{1} \rangle_i$, $| \widetilde{0} \rangle_i$, and $| -\widetilde{1} \rangle_i$ are the states of the $i$th nitroxide nucleus dressed by the hyperfine interaction with the nitroxide electron~\cite{SM}. Similarly, for a nitroxide hosting $^{15}$N, we find
\begin{align}
\begin{aligned}
H_{n_i} \approx \left[ E_{1/2}^i |\widetilde{1/2}\rangle\langle \widetilde{1/2}|_i + E_{-1/2}^i |-\widetilde{1/2}\rangle\langle -\widetilde{1/2}|_i \right] J_i^z. \label{eq:trace2}
\end{aligned}
\end{align}
There are now two energy-transition branches with transition energies given by $E_{1/2,-1/2}^i=\mu_B B^z G(\theta_i) \pm \frac{1}{2\sqrt{2}}\sqrt{(A^\perp)^2+(A^\parallel)^2+\left[(A^\parallel)^2-(A^\perp)^2\right]\cos(2\theta_i)}$~\cite{SM}. Fig.~\ref{fig:energies}(a) compares the energy-transition branches in Eqs.~(\ref{eq:trace1}) and (\ref{eq:trace2}) (dashed lines) with numerical diagonalization of Eq.~(\ref{eq:nitro}) (solid lines). Note that since we will target the $E_0^i$ branch for label detection, we have further developed its functional form to include second-order corrections in the electron-nitrogen coupling~\cite{SM}. Fig.~\ref{fig:energies}(a) focuses on a single nitroxide (the index $i=1$ is removed for clarity) and shows that our expression for $E_0$ is in excellent agreement with numerical diagonalization.

The inter-label coupling leads to an additional splitting $\propto g_{12}$ of the energy-transition branches in Eqs.~(\ref{eq:trace1}, \ref{eq:trace2}). This term typically simplifies as $g_{12}\left[J_1^z J_2^z-\frac{1}{4}\left(J_1^+ J_2^- + J_1^- J_2^+ \right)\right]\longrightarrow g_{12} J_1^z J_2^z$, since labels with different orientations have different resonance frequencies, thus eliminating the flip-flop contribution. Note that while this simplification is frequently (but not always) valid when using two $^{14}$N, it is guaranteed to be valid when using nitroxides hosting different nitrogen isotopes since $E_0$ (for $^{14}$N) differs from $E_{1/2,-1/2}$ (for $^{15}$N) for any value of the azimuth [see Fig.~\ref{fig:energies}(a)]. For $^{14}$N, the available resonances of, e.g., the first label are then determined by the modified nitroxide Hamiltonian $H'_{n_1} = H_{n_1}+g_{12} J_1^z J_2^z$
\begin{align}
\begin{aligned}
H'_{n_1} &= \left[ E_1^1 |\widetilde{1}\rangle\langle \widetilde{1}|_1 + E_0^1 |\widetilde{0}\rangle\langle \widetilde{0}|_1 + E_{-1}^1 |-\widetilde 1\rangle\langle -\widetilde{1}|_1 \right] J_1^z + g_{12} J_1^z J_2^z \\
&=\sum_{q=1,0,-1}\left[\left(E_q^1 + \frac{g_{12}}{2}\right) |\widetilde{q} e\rangle\langle \widetilde{q} e| + \left(E_q^1 - \frac{g_{12}}{2}\right) |\widetilde{q} g\rangle\langle \widetilde{q} g|\right] J_1^z.
\end{aligned}
\end{align}
That is, any energy-transition branch $E_q^1$ (corresponding to the $|\widetilde{q}\rangle$ nuclear state of the first nitroxide) splits as
\begin{align}
\begin{aligned}
E_q^1 \longrightarrow \left(E_q^1 + \frac{g_{12}}{2}, E_q^1 - \frac{g_{12}}{2}\right) \label{eq:interLabelSplitting}
\end{aligned}
\end{align}
depending on the electronic state $|e\rangle$, $|g\rangle$ of the second nitroxide. Ref.~\cite{SM} includes a complete energy diagram of an electron spin in a nitroxide.

\subsection*{The Protocol}
Our DEER protocol simultaneously delivers MW and RF pulses to detect energy shifts in the NV spectrum due to inter-label coupling. The sequence contains two free-evolution stages of duration $\tau_{\rm free}$ separated by a driving stage [see Fig.~\ref{fig:system}(c)]. The NV is prepared in the $+1$ eigenstate of $\sigma^x$ and the nitroxides are assumed to be in a fully thermalized state. During free evolution, the NV accumulates a phase due to the term $\sigma^z \left(\vec{a}_1 \cdot \vec{J}_1+\vec{a}_2 \cdot \vec{J}_2\right)$ in Eq.~\eqref{eq:main}. This term can be approximated as $\sigma^z \left(a_1^z J_1^z+a_2^z J_2^z\right)$ in the rotating-wave approximation (RWA) since the $J_i^{x,y}$ contributions are suppressed by the large electronic precession frequencies of the nitroxides. Indeed, the energies $E_{1,0,-1}^i$ reach several hundreds of MHz even for moderate values of $B^z$ (we use $B^z = 30$\ mT in our simulations). By contrast, the NV-label coupling takes values of hundreds of kHz for typical NV-label distances of several nanometers.
\begin{figure}[t]
\hspace{-0.5cm}\includegraphics[width=1.05\columnwidth]{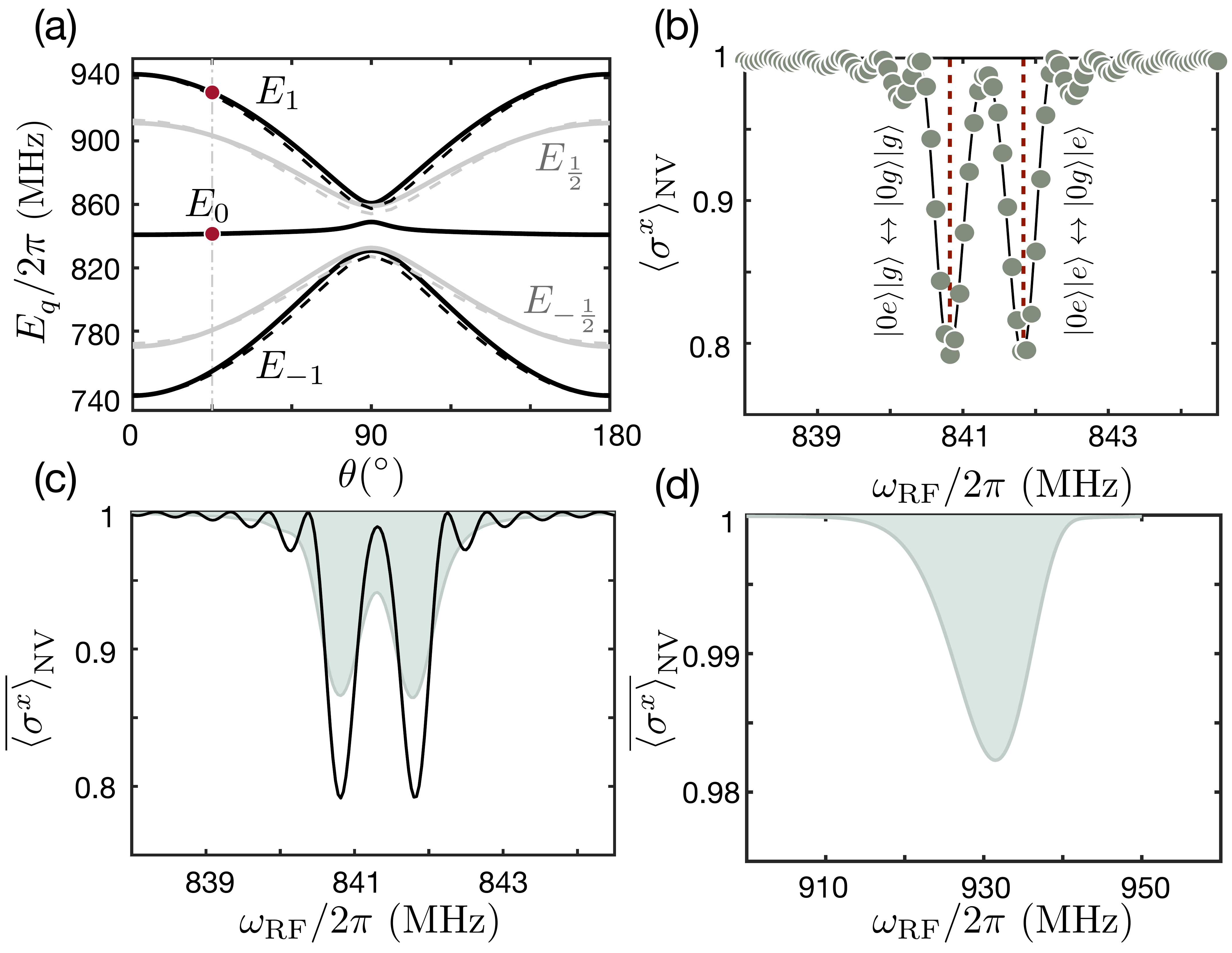}
\caption{(a) Energy-transition branches of an electron spin label at $B^z = 30$\ mT for $^{14}$N ($^{15}$N). Solid black (grey) lines show the transition energies as a function of the azimuth $\theta$ obtained via numerical diagonalization of Eq.~(\ref{eq:nitro}). Dashed black (grey) lines correspond to $E_{1,0,-1}$ in Eq.~(\ref{eq:trace1}) [$E_{1/2,-1/2}$ in Eq.~(\ref{eq:trace2})]. The vertical line highlights the case $\theta_i = 30^\circ$, with red dots indicating the transition energies relevant in the following plots. (b) NV spectrum $\langle\sigma^x\rangle_{\rm NV}$ as a function of the RF frequency applied near $E_0^1$. The spectrum was obtained by unitary propagation of Eq.~\eqref{eq:main} with $H_{n_i}$ given by Eq.~\eqref{eq:nitro} (black line) and by $E_0^i J_i^z$ (grey circles). The first nitroxide has azimuth $\theta_1 = 30^\circ$ while the second nitroxide has azimuth $\theta_2 = 91.7^\circ$. The splitting of the peaks is induced by the inter-label coupling $g_{12}\approx 2\pi\times 1$ MHz. The positions of the resonances are indicated by the dashed red lines. In (c, d), shaded areas show the average NV spectra $\overline{\langle\sigma^x\rangle}_{\rm NV}$ in the presence of tumbling of the first (near-resonant) nitroxide. Tumbling was mimicked by averaging over a random azimuth following a Gaussian distribution centered at the equilibrium value $\theta_{1,{\rm eq}}=30^\circ$. The orientation of the other nitroxide was kept fixed for simplicity. The standard deviation was chosen to be $\sigma_\theta = 6.25^\circ$. In (c), we observe that tumbling leads to a loss of contrast and to signal broadening near $E_0^1$. Nevertheless, the spectrum still shows two clearly separated peaks. For comparison, the case without tumbling in (b) is superimposed (black line). (d) Average NV spectrum $\overline{\langle\sigma^x\rangle}_{\rm NV}$ near $E_1^1$. Here, the energy splitting cannot be resolved due to nitroxide tumbling. Also note that tumbling severely reduces contrast. For that reason, the spectrum in the absence of tumbling has much higher contrast and is not shown. \label{fig:energies}}
\end{figure}
During MW/RF irradiation, the NV undergoes an overall flip $\sigma^z \rightarrow -\sigma^z$ via $2N+1$ contiguous $\pi$-pulses arising from the same continuous MW drive. At the same time, a single weaker RF $\pi$-pulse on the labels induces $J_i^z\rightarrow-J_i^z$ if it is near-resonant with an electronic transition (i.e., with $E_{1,0,-1}^i$ for $^{14}$N or with $E_{1/2,-1/2}^i$ for $^{15}$N). The simultaneous flipping of the NV and spin label ensures a constructive and coherent accumulation of the NV phase during free evolution. Meanwhile, the use of $2N+1$ $\pi$-pulses on the NV ensures that the Rabi frequencies of the two drives are well separated, which effectively averages out spurious NV-label interactions during irradiation. Consequently, scanning the RF frequency near nitroxide transitions yields clean resonance peaks in the NV response. As long as $\Omega_{\rm RF}< g_{12}$, the resonance peaks are split by $g_{12}$. Fig.~\ref{fig:energies}(b) shows the NV spectrum (the expectation value of $\sigma^x$) associated with the transitions $|0g\rangle |g\rangle \leftrightarrow |0e\rangle |g\rangle$
and $|0g\rangle |e\rangle \leftrightarrow |0e\rangle |e\rangle$. The resonances are split by $\sim g_{12}$ as expected. The solid black line in Fig.~\ref{fig:energies}(b) was obtained by numerically propagating Eq.~\eqref{eq:main} with $H_{n_i}$ given by Eq.~\eqref{eq:nitro}, while grey circles were obtained assuming the simplified $H_{n_i}$ in Eq.~\eqref{eq:trace1}. The two spectra are in excellent agreement. Simulations in Fig.~\ref{fig:energies}(b) assume $B^z = 30$\ mT, which is in the range of values that ensure the stability of the $E_0^i$ energy-transition branch discussed below~\cite{SM}. Moreover, the parameters of the pulse sequence are $\tau_{\rm free} = 1.3\ \mu{\rm s}$, $\Omega _{\rm RF} = 2\pi \times 250$\ kHz, and $\Omega_{\rm MW} = 31\times \Omega_{\rm RF}$, for a total sequence time of $4.6\ \mu$s. In addition, the labels are separated from the NV by the distances $d_1 \approx 6.9$ nm and $d_2 \approx 7.3$ nm, leading to $\vec{a}_1 \approx 2\pi \times \left(128, -132, -223\right)$ kHz and $\vec{a}_2 \approx 2\pi \times \left(-22, -16, -264\right)$ kHz, while the distance between labels is $d_{12} \approx 3.297$ nm.

The expressions for the energy-transition branches in Eqs.~(\ref{eq:trace1}, \ref{eq:trace2}) reveal a dependence on the azimuth angle, i.e., $E_q^i\equiv E_q^i(\theta_i)$. Consequently, unavoidable protein motion during the irradiation stage leads to a distortion in the spectrum and to a difficult interpretation of the signal. However, it is important to note the distinct nature of the $E_0^i$ branch, which shows a much weaker dependence on $\theta_i$ than $E_{1,-1}^i$. This makes the $E_0^i$ branch particularly well suited for robust detection of the energy splitting [see Figs.~\ref{fig:energies}(c,d)]. To maximize the robustness, we chose the magnetic field to be large enough to energetically suppress the effect of the anisotropic hyperfine interaction, but small enough that the anisotropy of the Land{\'e} tensor does not become significant~\cite{SM}.

\subsection*{Numerical simulations}
\begin{figure}[t]
\hspace{-0.5 cm}\includegraphics[width=1.05\columnwidth]{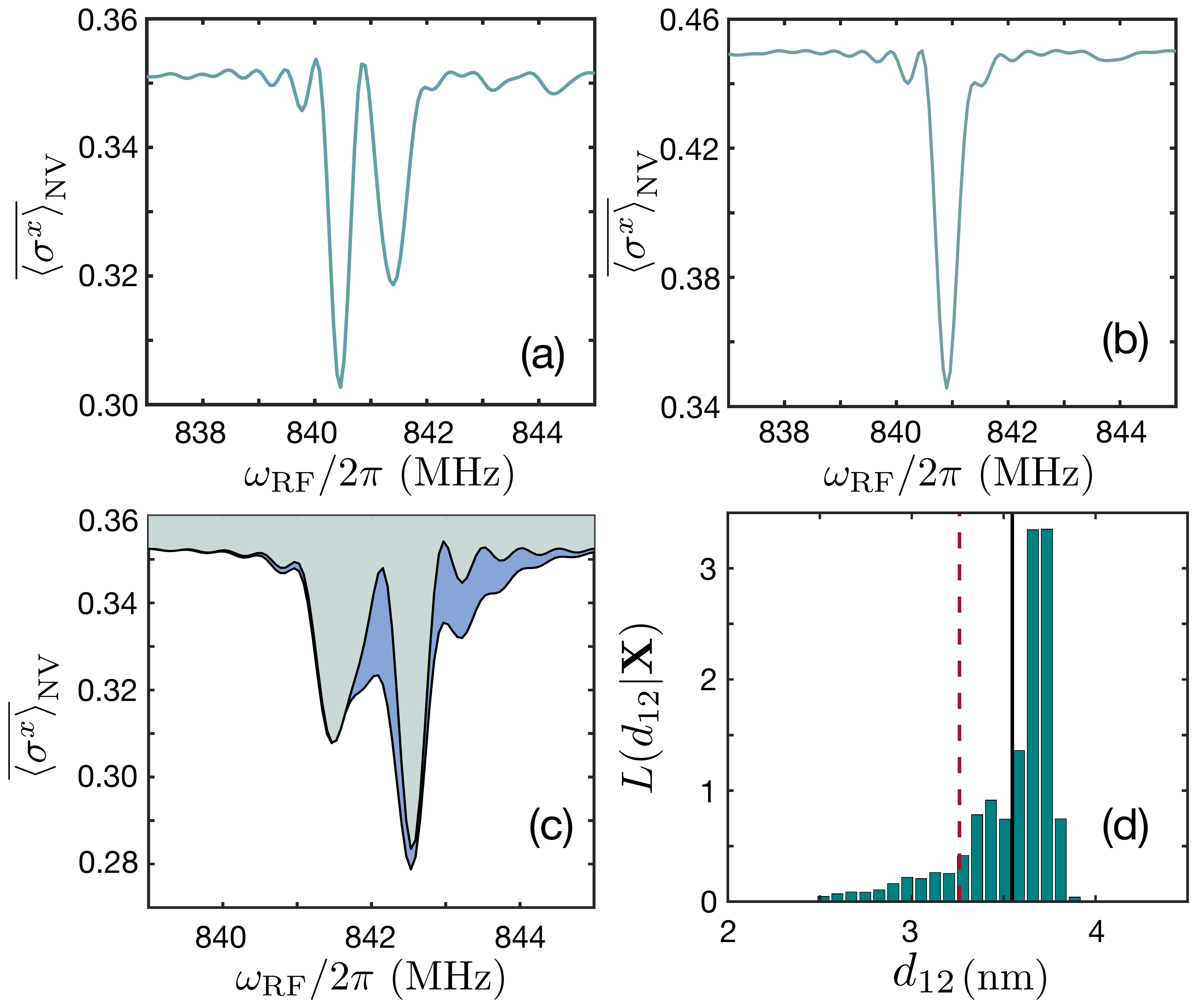}
\caption{(a) Simulated average NV spectrum $\overline{\langle\sigma^x\rangle}_{\rm NV}$. Two resonance peaks due to $g_{12}$ are observed. (b) Similar to (a) but with lower $g_{12}$. (c) Simulated average spectrum $\overline{\langle\sigma^x\rangle}_{\rm NV}$ for overlapping label resonances with two $^{14}$N isotopes (blue shaded area) and with distinct $^{14}$N and $^{15}$N isotopes (grey shaded area). (d) Marginal distribution $L(d_{12}|{\bf X})$ of the inter-label distance for simulated data acquisition using the spectrum shown in (a). The ``true'' value of the distance and its posterior expectation, $3.297$ nm and $3.54(25)$ nm, are indicated with vertical lines. The error is the standard deviation of the marginal posterior. \label{fig:tilting}}
\end{figure}

We now illustrate our method in realistic ambient conditions including decoherence leading to NV dephasing with decoherence time $T_2 = 20\ {\rm\mu s}$ (for a 4 nm NV depth~\cite{Shi15}), electron-spin label relaxation with $T_1 = 4\ {\rm \mu s}$~\cite{Shi15}, and molecular tumbling. The dissipative model used in this section is described in Ref.~\cite{SM} and captures the main decoherence mechanisms identified in Ref.~\cite{Shi15}. Note that for the specific protocol considered here, decoherence mainly manifests as a reduced line contrast and not as an increased linewidth. This is because we keep the sequence duration fixed and sweep $\omega_{\rm RF}$ to find the resonances, as opposed to varying the sequence length at fixed $\omega_{\rm RF}$ to observe an echo signal. The molecular tumbling is modelled as a random rigid rotation of both nitroxides around an axis parallel to the laboratory $x$ axis~\cite{SM}. Our simulations use a rotation angle $\delta$ that is Gaussian-distributed with standard deviation $\sigma_\delta = 6.25^\circ$. This is somewhat smaller but comparable to the fluctuations observed in Ref.~\cite{Shi15}. Note that immobilizing proteins in more rigid matrices~\cite{Meyer15,Lira16} or attaching them to the diamond surface~\cite{Lovchinsky16} through multiple rigid covalent bonds could further reduce tumbling. The resulting tumbling-averaged spectra $\overline{\langle\sigma^x\rangle}_{\rm NV}$ are shown in Fig.~\ref{fig:tilting}. In Figs.~\ref{fig:tilting}(a,b,c), the parameters $\tau_{\rm free}$, $\Omega_{\rm MW}$, $\Omega_{\rm RF}$, and $\vec{a}_i$ are the same as in Fig.~\ref{fig:energies}. The details of the equilibrium nitroxide configurations are given in Ref.~\cite{SM}. In Fig.~\ref{fig:tilting}(a), the labels are separated by $d_{12} \approx 3.297$ nm and the inter-label coupling is $g_{12} \approx 2 \pi \times 1$ MHz. When the frequency of the RF field is set close to $E_0^1$, we clearly identify two resonance peaks in spite of tumbling. We have verified that the two resonances are still visible even if we choose $\sigma_\delta$ to be four times larger. Since $g_{12}\propto d_{12}^{-3}$, a change in the relative position of the labels significantly modifies the observed spectrum. This is shown in Fig.~\ref{fig:tilting}(b) where the second label was displaced such that $d_{12} \approx 4.03$ nm, leading to the disappearance of the splitting. This change in the spectrum certifies a change in the relative position between labels and, by extension, a conformational change in the host molecule.

So far, all simulations were performed for nitroxides hosting $^{14}$N and having distinct transition energies. If both nitroxides have similar transition energies (e.g., due to similar azimuths), the spectra of the two nitroxides overlap and the inter-label flip-flop terms cannot be neglected. This complicates the interpretation of the spectrum. As shown in Fig.~\ref{fig:tilting}(c), this problem is avoided by using distinct $^{14}$N and $^{15}$N isotopes in each nitroxide. In this case, the $E_0$ branch of the $^{14}$N nitroxide never overlaps with the branches of the $^{15}$N nitroxide and flip-flop terms can be safely neglected. As a result, the interpretation of the spectrum is much simpler. We emphasize that the purpose of using such ``orthogonal labels'' is not merely to resolve their distinct spectral signatures~\cite{Galazzo22}: rather, it is to simplify the form of the inter-label interaction itself for all label orientations.

\subsection*{Bayesian inference}
Finally, we show how to infer the inter-label distance $d_{12}$ under realistic ambient conditions. We first simulate the experimental acquisition of $\overline{\langle\sigma^x \rangle}_{\rm NV}$. The simulated dataset has the form ${\bf X}=\{ X_1,\ldots,X_M\}$, where $X_j$ is an experimental estimate of the probability of measuring $\sigma^x = + 1$ after $N_m$ measurements at RF frequency $\omega_{{\rm RF},j}$. Second, we use a simplified model to efficiently extract information from ${\bf X}$. The model captures the relevant features of the tumbling-averaged spectrum $\overline{\mathcal{S}}(\omega_{\rm RF})$. More precisely, we derive the approximate expression $\mathcal{S}(\omega_{\rm RF},\theta,\beta)=\mathcal{S}_0 - \sum_{s=+,-} \mathcal{C}_s \left[\frac{\Omega_{\rm RF}}{\Omega_s(\omega_{\rm RF},\theta,\beta)}\right]^2 \sin^2\left[\frac{\pi}{2}\frac{\Omega_s(\omega_{\rm RF},\theta,\beta)}{\Omega_{\rm RF}}\right]$ for a specific nitroxide azimuth $\theta$ and a specific angle $\beta$ between the magnetic field and the line joining the nitroxides. Here, $\Omega_\pm^2(\omega_{\rm RF},\theta,\beta) = \Omega_{\rm RF}^2 + \left[\omega_{\rm RF}-E_0(\theta) \pm g_{12}(\beta)/2 \right]^2$ and $\mathcal{S}_0$ and $\mathcal{C}_\pm$ are parameters that adjust the baseline and contrast, respectively~\cite{SM}. Moreover, $E_0(\theta)$ is the ${}^{14}$N energy-transition branch, while $g_{12}(\beta)\propto d_{12}^{-3}(1-3\cos^2\beta)$ is the inter-label coupling. Both $\theta$ and $\beta$ depend on the tumbling angle $\delta$~\cite{SM}. Averaging $\mathcal{S}[\omega_{\rm RF},\theta(\delta),\beta(\delta)]$ over a Gaussian distribution for $\delta$ gives the tumbling-averaged spectrum $\overline{\mathcal{S}}(\omega_{\rm RF})$. Assuming that the baseline $\mathcal{S}_0$ is known, our model contains eight free parameters denoted by ${\bf V}$~\cite{SM}. These include $d_{12}$ and the unknown standard deviation $\sigma_\delta$ of the tumbling. It must be emphasized that the light tumbling dependence of the $E_0$ branch enables the determination of the inter-label distance. Indeed, a simple measurement of the line splitting $g_{12} \propto d_{12}^{-3}(1-3\cos^2\beta)$ cannot give independent access to the distance $d_{12}$ and the angle $\beta$. This is because $\beta$ is unknown {\it a priori}. However, light tumbling fluctuations allow the NV to probe different geometric configurations of the nitroxides. This results in a small distortion of the line shape that yields information beyond that contained in the splitting $g_{12}$. This in turn enables the independent extraction of the distance $d_{12}$ and of the angle $\beta$. With the help of our model, we can therefore obtain the posterior probability of the parameters ${\bf V}$ using Markov Chain Monte Carlo sampling~\cite{Gilks96}. Assuming a uniform prior for ${\bf V}$, the posterior probability for ${\bf V}$ is $L({\bf V}|{\bf X})=\prod_{j=1,M}e^{-[X_j-(\overline{\mathcal{S}}(\omega_{{\rm RF},j})+1)/2]^2/2\sigma_m^2}/\sqrt{2\pi \sigma_m^2}$. Here, the noise variance $\sigma_m^2 \approx \left(1-\overline{\langle\sigma^x\rangle}_{\rm NV}^2\right)/4 N_m$ is assumed to be approximately constant and known. In Fig.~\ref{fig:tilting}(d) we show the resulting marginal $L(d_{12}|{\bf X})$ of $d_{12}$ for a dataset ${\bf X}$ simulated from the spectrum shown in Fig.~\ref{fig:tilting}(a). Here, ${\bf X}$ was obtained by taking $N_m = 2\times 10^4$ ideal measurements for each of $M=25$ frequencies ranging from $\omega_{\rm RF}/2\pi=839$ to $843$ MHz (we estimate that the same accuracy is achieved with $N_m \approx 5\times10^5$ for imperfect NV detection efficiency~\cite{SM,Wan18}). The expectation of the marginal posterior is $d_{12}=3.54(25)$ nm, close to the ``real'' value $3.297$ nm [see Fig.~\ref{fig:tilting}(d)].


\section*{DISCUSSION}
Our results open many interesting avenues for future investigation. In particular, it would be of great interest to extend our scheme to molecules with more than two attached spin labels. This would yield a stronger NV response, leading to a more detailed observation of conformational changes as well as to an enhancement in the range of inter-label distances to be estimated. The scheme could also be improved through the use of multiple NVs to better triangulate the label positions. In addition, our analytical understanding of the NV response could enable fast Bayesian inference of molecular dynamical properties. Our results pave the way for the observation of conformational dynamics of individual proteins using tools from magnetic resonance.

\section*{CODE AVAILABILITY}
All the codes employed are available upon request to the authors.

\section*{ACKNOWLEDGEMENTS}
\emph{Acknowledgements.--} The authors acknowledge financial support from Spanish Government via PGC2018-095113-B-I00 (MCIU/AEI/FEDER, UE) and, from Basque Government via IT986-16. C.M.-J. acknowledges the predoctoral MICINN grant PRE2019-088519. R.P. acknowledges support from European Union's Horizon 2020 FET-Open project SuperQuLAN (899354). M.B.P. and B.D. acknowledge the ERC Synergy Grants HyperQ (856432), as well as the BMBF project QSens (03ZU1110FF), and the EU project Asteriqs (820394). The authors acknowledge support by the state of Baden-W\"urttemberg through bwHPC and the German Research Foundation (DFG) through grant no INST 40/575-1 FUGG (JUSTUS 2 cluster). J.~C. acknowledges the Ram\'{o}n y Cajal   (RYC2018-025197-I) research fellowship, the financial support from Spanish Government via EUR2020-112117 and Nanoscale NMR and complex systems (PID2021-126694NB-C21) projects, the EU FET Open Grant Quromorphic (828826),  the ELKARTEK project Dispositivos en Tecnolog\'i{a}s Cu\'{a}nticas (KK-2022/00062), and the Basque Government grant IT1470-22.

${}^*$ carlosmunueraj@gmail.com

 \section*{AUTHOR CONTRIBUTIONS}
 All authors contributed to the main ideas and to the manuscript writing; C.M.-J. developed the analytical study; C.M.-J., R.P. and B.D. produced the numerical simulations; M.B.P. and J.C. led the work jointly.  

\bibliographystyle{naturemag}
\bibliography{article}

\clearpage
\widetext
\begin{center}
\textbf{ \large Supplemental Material: \\ Detection of Molecular Changes with Nitrogen-Vacancy Centers and Electron-Spin Labels}
\end{center}
\setcounter{equation}{0} \setcounter{figure}{0} \setcounter{table}{0}
\setcounter{page}{1} \makeatletter \global\long\def\theequation{S\arabic{equation}}
 \global\long\def\thefigure{S\arabic{figure}}
 \global\long\def\bibnumfmt#1{[S#1]}
 \global\long\def\citenumfont#1{S#1}

\section{NV-labels Hamiltonian}
In the following, we derive Eq.~(1) of the main text. We start with the Hamiltonian for a driven NV center interacting with two driven nitroxide electron-spin labels. The full Hamiltonian is
\begin{align}
\begin{aligned}
H = D \left(S^z\right)^2 +B^z|\gamma_e| S^z + H_{n_1}+H_{n_2} +\sum_{i=1,2} H_{{\rm NV}-n_i} + H_{ee} + \sqrt{2}\Omega_{\rm MW}S^x\cos(\omega_{\rm MW}t) + 2 \Omega_{\rm RF} \left(J_1^x+J_2^x\right) \cos(\omega_{\rm RF} t). \label{eq:SfullHamiltonian}
\end{aligned}
\end{align}
Here, $D = 2\pi \times 2.87$ GHz is the zero-field splitting of the NV center, $|\gamma_e| = 2\pi\times 28$\ MHz/mT is the electron gyromagnetic ratio, $H_{{\rm NV}-n_i} = \frac{\mu_0 \gamma_e^2 \hbar}{4 \pi d_i^3}\left[\vec{S}\cdot\vec{J}_i-\frac{3\left(\vec{S}\cdot\vec{r}_i\right)\left(\vec{J}_i\cdot\vec{r}_i\right)}{d_i^2}\right]$ is the dipolar interaction between the NV and the $i$th label electron, and $H_{ee} = \frac{\mu_0 \gamma_e^2 \hbar}{4 \pi d_{12}^3}\left[\vec{J}_1\cdot\vec{J}_2-\frac{3\left(\vec{J}_1\cdot\vec{r}_{12}\right)\left(\vec{J}_2\cdot\vec{r}_{12}\right)}{d_{12}^2}\right]$ is the dipolar interaction between label electrons. In these equations, $\vec{r}_i$ is the vector joining the NV with the $i$th label, $\vec{r}_{12} = \vec{r}_1 - \vec{r}_2$ is the relative vector that connects the two labels, $d_i = |\vec{r}_i|$, and $d_{12} = |\vec{r}_{12}|$. The last two terms on the right hand side of Eq.~(\ref{eq:SfullHamiltonian}) are the MW and RF driving terms, respectively.
We move to the interaction picture with respect to $H_0 = D \left(S^z\right)^2+|\gamma_e| B^z S^z = \omega_+ |1\rangle\langle 1 | + \omega_- |-1\rangle\langle -1|$, with $\omega_\pm = D\pm|\gamma_e|B^z$. In addition, we set the MW field on resonance with the $0\leftrightarrow 1$ NV transition ($\omega_{\rm MW} = \omega_+$). In the interaction picture, the Hamiltonian takes the form
\begin{align}
\begin{aligned}
H_I = H_{n_1} + H_{n_2} +S^z\left(\vec{a}_1 \cdot \vec{J}_1+\vec{a}_2 \cdot \vec{J}_2\right)+ H_{ee} + \frac{\Omega_{\rm MW}}{2}\left(|0\rangle\langle1|+|1\rangle\langle0|\right) + 2 \Omega_{\rm RF} \left(J_1^x+J_2^x\right) \cos(\omega_{\rm RF} t).
\end{aligned}
\end{align}
Note that we substituted $H_{{\rm NV}-n_i} \rightarrow S^z \vec{a}_i \cdot \vec{J}_i$, with $\vec{a}_i = \frac{\mu_0 \gamma_e^2 \hbar}{4 \pi d_i^3}\left[\hat{z}-\frac{3 r_i^z \vec{r}_i}{d_i^2}\right]$. This is justifiable since the strength of the interaction between the NV and labels is much smaller than the transition energies of the NV. Therefore, the terms proportional to $S^{x,y}$ can be neglected in the RWA.

We initialize the NV in the $m_s = 1,0$ manifold. Moreover, the Hamiltonian does not contain terms that can significantly populate the $m_s = -1$ subspace. Consequently, we can project the Hamiltonian on the $m_s = 1,0$ manifold and treat the NV as a two-level system. The resulting Hamiltonian is
\begin{align}
\begin{aligned}
H_I = H_{n_1} + H_{n_2} +\frac{\mathbb{I}+\sigma^z}{2}\left(\vec{a}_1 \cdot \vec{J}_1+\vec{a}_2 \cdot \vec{J}_2\right) +H_{ee} +\frac{\Omega_{\rm MW}}{2}\sigma^x + 2 \Omega_{\rm RF} \left(J_1^x+J_2^x\right) \cos(\omega_{\rm RF} t),
\end{aligned}
\end{align}
where $\mathbb{I} = |1\rangle\langle 1| + |0\rangle\langle 0|$, $\sigma^z = |1\rangle\langle 1| - |0\rangle\langle 0|$, and $\sigma^x = |1\rangle\langle 0| + |0\rangle\langle 1|$.

The next approximation consists in removing the counter-rotating terms in $H_{ee}$. These are terms of the form $J_1^+J_2^+$ and $J_1^-J_2^-$ that precess under the externally applied magnetic field $B^z$. They can be neglected in the RWA since the strength of the coupling between label electrons is much smaller than their Zeeman energy. Under this assumption, $H_{ee}$ simplifies to
\begin{align}
\begin{aligned}
H_{ee}\approx \frac{\mu_0 \gamma_e^2 \hbar}{4 \pi d_{12}^3}\left[1-3\left(\frac{r_{12}^z}{d_{12}}\right)^2\right]\left[J_1^z J_2^z-\frac{1}{4}\left(J_1^ + J_2^- + J_1^-J_2^+\right)\right],
\end{aligned}
\end{align}
with $J_i^\pm = J_i^x\pm i J_i^y$ and $g_{12} = \frac{\mu_0 \gamma_e^2 \hbar}{4 \pi d_{12}^3}\left[1-3\left(\frac{r_{12}^z}{d_{12}}\right)^2\right]$. Combining the above approximations leads to Eq.~(1) of the main text.

\section{Simplified dynamics of the nitroxide labels}
In this section, we develop a simplified model of electron-nucleus dynamics within each nitroxide label. Our approach is based on a perturbative treatment of the nitroxide nuclear degrees of freedom that are orthogonal to the $z$ direction, i.e., the direction of the external magnetic field. We first introduce the rotation matrices used to relate the laboratory axes and the nitroxide principal axes, namely,
\begin{align}
R^y(\theta) =
\begin{pmatrix}
\cos\theta & 0 & \sin\theta\\
0 & 1 & 0\\
-\sin\theta & 0 & \cos\theta
\end{pmatrix}, \quad
R^z(\varphi) =
\begin{pmatrix}
\cos\varphi & -\sin\varphi & 0\\
\sin\varphi & \cos\varphi & 0\\
0 & 0 & 1
\end{pmatrix}.
\end{align}
The two frames are related via the combined rotation $R(\theta,\varphi) = R^z(\varphi)R^y(\theta)$. In particular, a tensor $\mathbb{O}^{(P)}$ defined with respect to the nitroxide principal axes takes the form $\mathbb{O} = R(\theta,\varphi)\mathbb{O}^{(P)} R(\theta,\varphi)^{\intercal}$ in the laboratory frame. For instance, consider the nitroxide Hamiltonians [Eq.~(2) of the main text]. In the present analysis, we ignore the quadrupolar and nuclear Larmor terms since they are small and do not contribute significantly to the dynamics. The validity of this last assumption was confirmed by our detailed numerical simulations. With this assumption, the nitroxide Hamiltonians expressed in the laboratory frame become
\begin{align}
\begin{aligned}
H_{n_i} \approx \mu_B B^z \hat{z} \cdot \mathbb{G}_i \cdot \vec{J}_i+\vec{J}_i \cdot \mathbb{A}_i \cdot \vec{I}_i = \mu_B B^z \hat{z} \cdot R(\theta_i,\varphi_i) \mathbb{G}_i^{(P)} R(\theta_i,\varphi_i)^{\intercal} \cdot \vec{J}_i + \vec{J}_i \cdot R(\theta_i,\varphi_i)\mathbb{A}_i^{(P)} R(\theta_i,\varphi_i)^{\intercal} \cdot \vec{I}_i.
\end{aligned}
\end{align}
Due to the symmetry of the $\mathbb{G}_i^{(P)}$ and $\mathbb{A}_i^{(P)}$ tensors ($G^x\approx G^y\approx 2.007$ and $A^x\approx A^y\approx 2\pi \times 14 \ {\rm MHz}$) we can set $\varphi_i = 0$ without loss of generality. We find
\begin{align}
\begin{aligned}
H_{n_i} = &\frac{\mu_B B^z}{2} (G^\parallel - G^\perp) \sin (2 \theta_i) J_i^x + \frac{\mu_B B^z}{2} \left[( G^\parallel- G^\perp) \cos (2 \theta_i)+ G^\perp+ G^\parallel\right]J_i^z \\
&+\frac{A^{\parallel }- A^{\perp }}{2} \sin(2\theta_i)J_i^z I_i^x +\left(\frac{A^{\parallel } + A^{\perp}}{2} + \frac{A^{\parallel} - A^{\perp}}{2} \cos(2\theta_i)\right)J_i^z I_i^z \\
&+A^{\perp} J_i^y I_i^y + \left(\frac{A^{\parallel } + A^{\perp}}{2} - \frac{A^{\parallel} - A^{\perp}}{2} \cos(2\theta_i) \right)J_i^x I_i^x + \frac{A^{\parallel}- A^{\perp}}{2} \sin(2\theta_i)J_i^x I_i^z. \label{eq:Sroot}
\end{aligned}
\end{align}

\subsection{Simplified dynamics of a nitroxide label hosting $^{14}$N}
\begin{figure}[h]
\hspace{-0. cm}\includegraphics[width=0.5\columnwidth]{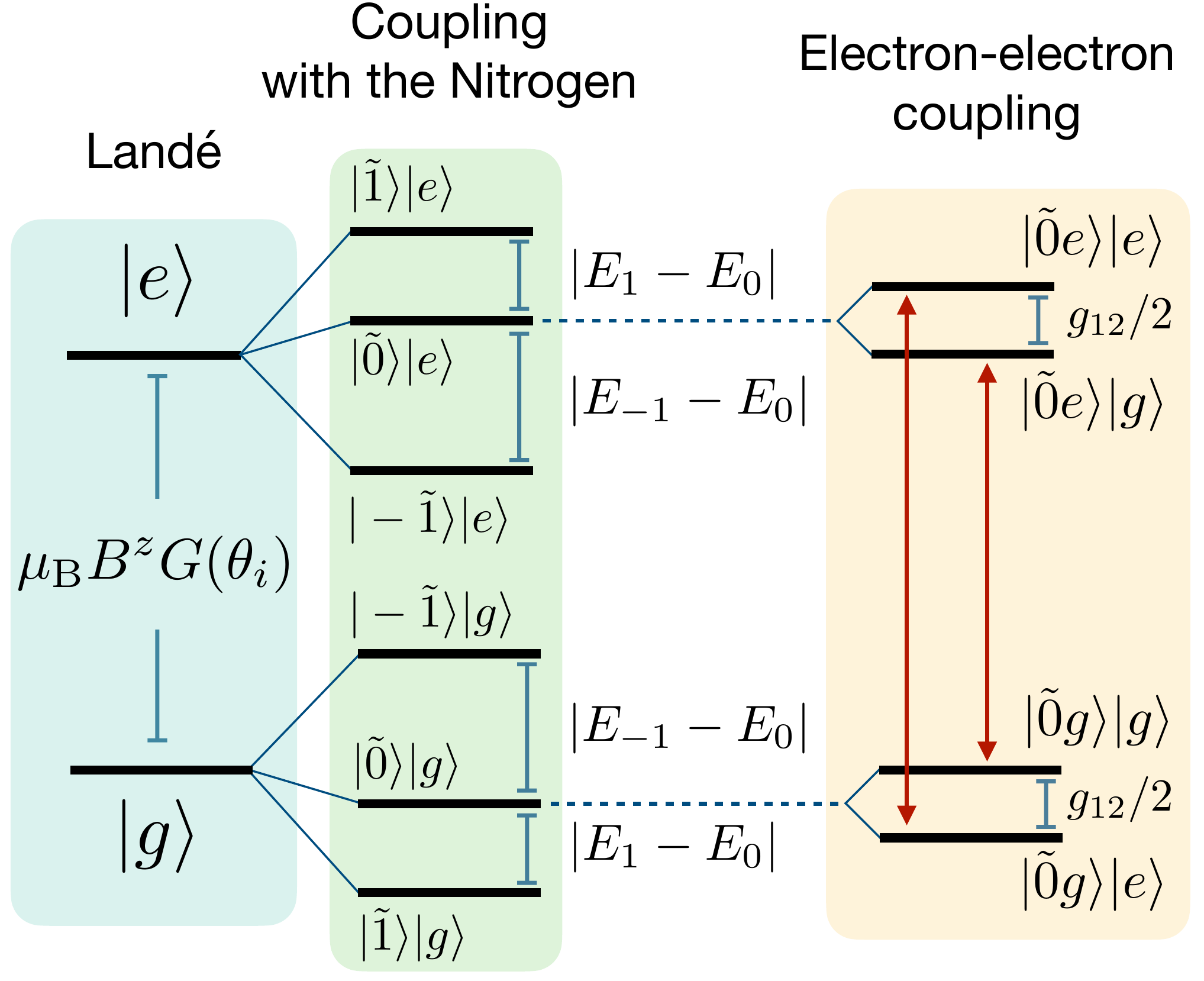}
\caption{Energy diagram of a nitroxide label hosting a $^{14}$N. A magnetic field $B^z$ is applied along the laboratory $z$ axis, leading to a Land\'e splitting of the electronic states (blue panel). The hyperfine coupling between the label electron and the nitrogen then splits the energy levels (green panel). Finally, each level is further split in two by the coupling between label electrons (yellow panel). The red arrows in the yellow panel indicate the transitions targeted by our protocol. \label{fig:Ssplittings}}
\end{figure}

\begin{figure}[h]
\hspace{-0. cm}\includegraphics[width=0.8\columnwidth]{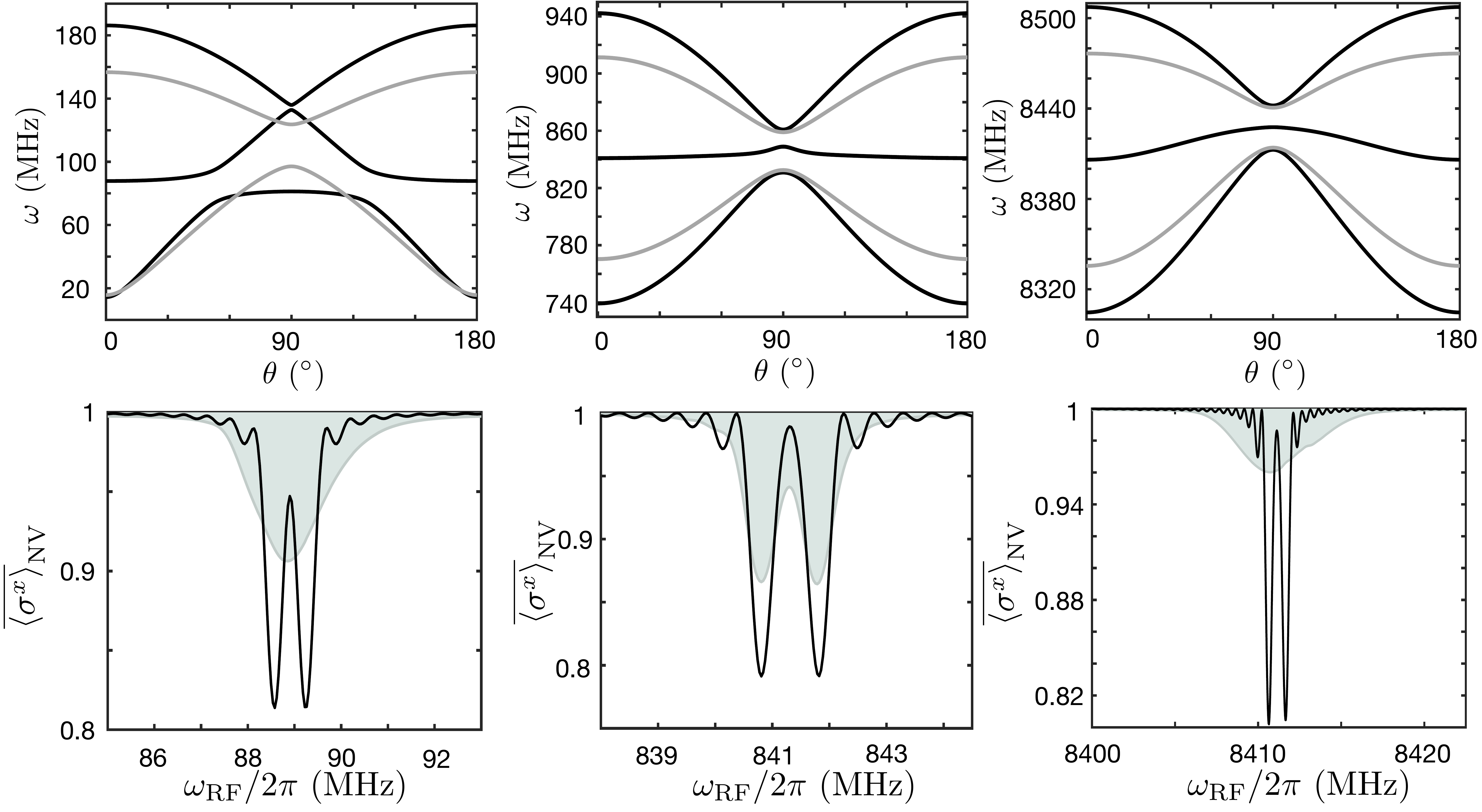}
\caption{(Top panels) Energy-transition branches of a nitroxide as a function of its azimuth angle $\theta$ for $B =$ $3$, $30$ and $300$\ mT (left to right). Solid black (grey) lines correspond to a nitroxide hosting $^{14}$N ($^{15}$N). All branches are obtained by diagonalizing Eq.~(2) of the main text. (Bottom panels) Numerical simulation of the average NV spectrum $\overline{\langle\sigma^x\rangle}_{\rm NV}$ for the same magnetic fields as in the top panels. All simulations are performed by unitary propagation of Eq.~(1) of the main text and using the full nitroxide Hamiltonian, Eq.~(2). We show the spectrum for an equilibrium azimuth $\theta_{\rm eq} = 30 ^\circ$ (black lines) and the spectra averaged over a Gaussian distribution of angles centered at $\theta_{\rm eq}$ and with standard deviation $\sigma_\theta = 6.25 ^\circ$ (shaded areas). At intermediate magnetic fields, the spectrum is resilient to tumbling due to a weak dependence of the central branch $E_0$ on $\theta$. \label{fig:SOrientation}}
\end{figure}

As a first approximation, we keep only the terms proportional to $J_i^z$ in Eq.~(\ref{eq:Sroot}) . This is a reasonable first approximation since the hyperfine interaction is typically smaller than the electronic Zeeman energy. The resulting Hamiltonian is
\begin{align}
\begin{aligned}
H_{\rm diag} &= \frac{\mu_B B^z}{2} \left[( G^{\parallel}- G^{\perp}) \cos (2 \theta_i )+ G^{\perp}+ G^{\parallel}\right]J_i^z +\left(\frac{A^{\parallel } + A^{\perp}}{2} + \frac{A^{\parallel} - A^{\perp}}{2} \cos(2\theta_i)\right) J_i^z I_i^z + \frac{A^{\parallel }- A^{\perp }}{2} \sin(2\theta_i)J_i^z I_i^x \\
&= \mu_B B^z G(\theta_i) J_i^z + \left[ \left(\frac{A^{\perp} + A^{\parallel}}{2} + \frac{A^{\parallel} - A^{\perp}}{2}\cos(2\theta_i) \right) I_i^z + \frac{A^{\parallel }- A^{\perp }}{2} \sin(2\theta_i)I_i^x \right]J_i^z, \label{eq:Sdiagonal}
\end{aligned}
\end{align}
where $G(\theta) =\frac{1}{2} \left[( G^\parallel- G^\perp) \cos (2 \theta )+ G^\perp+ G^\parallel\right]$. To find the shifts on $J_i^z$ associated to specific nuclear spin states, we diagonalize the term
\begin{align}
\begin{aligned}
\left(\frac{A^{\perp} + A^{\parallel}}{2} + \frac{A^{\parallel} - A^{\perp}}{2}\cos(2\theta_i) \right) I_i^z + \frac{A^{\parallel }- A^{\perp }}{2} \sin(2\theta_i)I_i^x. \label{eq:Stodiagonalise}
\end{aligned}
\end{align}
Since $^{14}$N is a spin-1 particle, diagonalization leads to three nuclear eigenstates $|\widetilde{0}\rangle$, $|\widetilde{1}\rangle$, and $|-\widetilde{1}\rangle$ . The states have eigenenergies
\begin{align}
\begin{aligned}
&\omega_0^i=0,\\
&\omega_1^i=\frac{1}{\sqrt{2}} \sqrt{\left[(A^{\parallel})^2- (A^{\perp})^2\right] \cos (2 \theta_i )+(A^{\perp})^2+(A^{\parallel})^2}, \\
&\omega_{-1}^i=-\frac{1}{\sqrt{2}} \sqrt{\left[(A^{\parallel})^2- (A^{\perp})^2\right] \cos (2 \theta_i )+(A^{\perp})^2+(A^{\parallel})^2}.
\end{aligned}
\end{align}
Thus, our leading approximation for the nitroxide Hamiltonian is
\begin{align}
\begin{aligned}
H_{n_i} \approx \left[E_1^i |\widetilde{1}\rangle\langle \widetilde{1}| + E_0^i |\widetilde{0}\rangle\langle \widetilde{0}| + E_{-1}^i |-\widetilde 1\rangle\langle -\widetilde{1}|\right] J_i^z, \label{eq:firstOrder}
\end{aligned}
\end{align}
where
\begin{align}
\begin{aligned}
&E_0^i= \mu_B B^z G(\theta_i), \\
&E_1^i= \mu_B B^z G(\theta_i) + \frac{1}{\sqrt{2}} \sqrt{\left[(A^{\parallel})^2- (A^{\perp})^2\right] \cos (2 \theta_i )+(A^{\perp})^2+(A^{\parallel})^2}, \\
&E_{-1}^i=\mu_B B^z G(\theta_i)-\frac{1}{\sqrt{2}} \sqrt{\left[(A^{\parallel})^2- (A^{\perp})^2\right] \cos (2 \theta_i )+(A^{\perp})^2+(A^{\parallel})^2}.
\end{aligned}
\end{align}
The above expressions reveal the existence of three energy-transition branches. There is one central branch with energy $\mu_B B^z G(\theta_i)$. The other two branches are shifted from the central branch by $\pm \left|\frac{1}{\sqrt{2}} \sqrt{\left[(A^{\parallel})^2- (A^{\perp})^2\right] \cos (2 \theta_i )+(A^{\perp})^2+(A^{\parallel})^2}\right|$. A full energy diagram is given in Fig.~\ref{fig:Ssplittings}.
Note that for moderate magnetic fields, $E_0^i$ depends much more weakly than $E_{1,-1}^i$ on the nitroxide azimuth $\theta_i$ since the Land{\'e} tensor is almost isotropic, $G^\parallel \simeq G^\perp$, while the hyperfine tensor is strongly anisotropic, $A^\parallel \neq A^\perp$. Thus, the central branch is particularly important since it can be robust against molecular tumbling (i.e., robust against changes in $\theta_i$). Note, however, that the dependence of $E_0^i$ on $\theta_i$ due to the small anistropy of the Land{\'e} tensor can become significant if the magnetic field becomes too large. At low magnetic fields, the above approximation starts to break down. We now investigate the leading correction to the central branch $E_0^i$ due to the hyperfine interaction in that regime. This correction arises from the nondiagonal terms in Eq.~\eqref{eq:Sroot}, i.e., the terms proportional to $J_i^{x,y}$. We first rewrite the nondiagonal terms in Eq.~\eqref{eq:Sroot} as
\begin{align}
\begin{aligned}
V = \alpha_i J_i^x+\beta_i J_i^y,
\end{aligned}
\end{align}
where
\begin{align}
\begin{aligned}
\alpha_i = \left(A^\perp\cos^2(\theta_i)+A^\parallel\sin^2(\theta_i)\right)I^x+\left(A^\parallel-A^\perp\right)I^z\cos(\theta_i)\sin(\theta_i), \quad \beta_i = A^\perp I^y.
\end{aligned}
\end{align}
We find the second-order correction due to $V$ by performing operator perturbation theory. More precisely, we move to a rotating frame with respect to $\mu_B B^z G(\theta_i) J_i^z$ and find the effective contribution of $V$ by keeping time-independent terms in its Dyson series (expanded up to second order). This leads to an effective interaction $\frac{(\alpha_i-i \beta_i)(\alpha_i+i\beta_i)}{2\mu_B B^z G(\theta_i)} J_i^z$. Projecting this expression onto the $|\widetilde{0}\rangle$ subspace gives the leading correction to $E_0^i$, 
\begin{align}
\begin{aligned}
\frac{1}{2 \mu_B B^z G(\theta_i)}\frac{2 (A^{\perp} A^{\parallel})^2 +\left[(A^{\perp})^4-(A^{\perp} A^{\parallel})^2\right] \sin^2(\theta_i)}{(A^{\perp})^2 \sin^2(\theta_i)+(A^{\parallel})^2 \cos^2(\theta_i)}.
\end{aligned}
\end{align}
This expression shows that the anisotropy of the hyperfine tensor can lead to a strong dependence of $E_0^i$ on $\theta_i$ when the magnetic field becomes small enough to be comparable to the hyperfine coupling. From the above arguments, we therefore expect that there exists an optimum magnetic field strength where the central branch is most robust to variations in the azimuth $\theta_i$. This is illustrated in Fig.~\ref{fig:SOrientation}.

In summary, we write the nitroxide Hamiltonian as
\begin{align}
\begin{aligned}
H_{n_i} \approx \left[ E_1^i |\widetilde{1}\rangle\langle \widetilde{1}|_i + E_0^i |\widetilde{0}\rangle\langle \widetilde{0}|_i + E_{-1}^i |-\widetilde 1\rangle\langle -\widetilde{1}|_i \right] J_i^z,
\end{aligned}
\end{align}
with
\begin{align}
\begin{aligned}
&E_0^i= \mu_B B^z G(\theta_i) + \frac{1}{2 \mu_B B^z G(\theta_i) }\frac{2 (A^{\perp} A^{\parallel})^2 +\left[(A^{\perp})^4-(A^{\perp} A^{\parallel})^2\right] \sin^2(\theta_i)}{(A^{\perp})^2 \sin^2(\theta_i)+(A^{\parallel})^2 \cos^2(\theta_i)}
, \\
&E_1^i= \mu_B B^z G(\theta_i) + \frac{1}{\sqrt{2}} \sqrt{\left[(A^{\parallel})^2- (A^{\perp})^2\right] \cos (2 \theta_i )+(A^{\perp})^2+(A^{\parallel})^2}, \nonumber\\
&E_{-1}^i=\mu_B B^z G(\theta_i)-\frac{1}{\sqrt{2}} \sqrt{\left[(A^{\parallel})^2- (A^{\perp})^2\right] \cos (2 \theta_i )+(A^{\perp})^2+(A^{\parallel})^2}.
\end{aligned}
\end{align}

\subsection{Simplified dynamics of a nitroxide label hosting $^{15}$N}
For a $^{15}$N, (i.e., a spin-1/2 particle) the diagonalization of Eq.~\eqref{eq:Stodiagonalise} yields two nitrogen eigenstates $|\widetilde{1/2}\rangle$ and $|-\widetilde{1/2}\rangle$ with eigenergies
\begin{align}
\begin{aligned}
E_{1/2,-1/2}^i=\mu_B B^z G(\theta_i) \pm \frac{\sqrt{(A^\perp)^2+(A^\parallel)^2+\left[(A^\parallel)^2-(A^\perp)^2\right]\cos(2\theta_i)}}{2 \sqrt{2}}.
\end{aligned}
\end{align}

\section{Details of the numerical simulations}
\subsection{Dissipative model}
Our detailed numerical simulations account for decoherence in ambient conditions using a Lindblad master equation of the form
\begin{align}
\begin{aligned}
\dot{\rho} = -i\left[H, \rho\right] &+ \frac{1}{2T_2}\left(\sigma^z \rho \sigma^z - \rho\right) \\
&+ \sum_{i=1,2}\left[\Gamma (\bar{n} + 1) \left(J_i^- \rho J_i^+ - \frac{1}{2}J_i^+J_i^-\rho-\frac{1}{2}\rho J_i^+J_i^-\right) + \Gamma \bar{n} \left(J_i^+ \rho J_i^- - \frac{1}{2}J_i^-J_i^+\rho-\frac{1}{2}\rho J_i^-J_i^+\right)\right].
\end{aligned}
\end{align}
Here, $T_2 = 20\ \mu$s is the NV coherence time for an NV depth of $4$\ nm~\cite{Shi15}, $\Gamma \approx 2\pi \times 2.68$ Hz is the relaxation rate of the label electrons, and $\bar{n} = 1/\left[\exp\left(\hbar \gamma_e B^z/k_B T\right) - 1\right]$ is the thermal occupation of a bosonic thermal bath of temperature $T$ that generates electronic transitions. For the magnetic field $B^z = 30$\ mT and the temperature $T = 300$\ K used in our simulations, this corresponds to an electronic relaxation time $T_1 = 1/\left[(2\bar{n}+1)\Gamma\right] \approx 4\ \mu$s~\cite{Shi15}. The coherent part of the evolution in the presence of driving is described by the Hamiltonian
\begin{align}
\begin{aligned}
H = &\frac{\Omega_{\rm MW}}{2} \sigma^x + g_{12}\left[J_1^z J_2^z-\frac{1}{4}\left(J_1^+ J_2^- + J_1^- J_2^+ \right)\right] \\
&+ \sum_{i=1,2}\left[\frac{1}{2}\left(\mathbb{I} + \sigma^z \right) \vec{a}_i \cdot \vec{J}_i + \mu_B B^z \hat{z}\cdot \mathbb{G}_i \cdot\vec{J}_i+ \gamma_N B^z I_i^z+\vec{I}_i\cdot\mathbb{Q}_i\cdot\vec{I}_i+\vec{J}_i\cdot\mathbb{A}_i\cdot \vec{I}_i + 2 \Omega_{\rm RF} J_i^x \cos(\omega_{\rm RF} t)\right]. \label{eq:Slindblad}
\end{aligned}
\end{align}
The above model was previously used to accurately describe the experimental results of Ref.~\cite{Shi15}.

\subsection{Molecular tumbling}
\begin{figure}[h]
\hspace{-0. cm}\includegraphics[width=0.8\columnwidth]{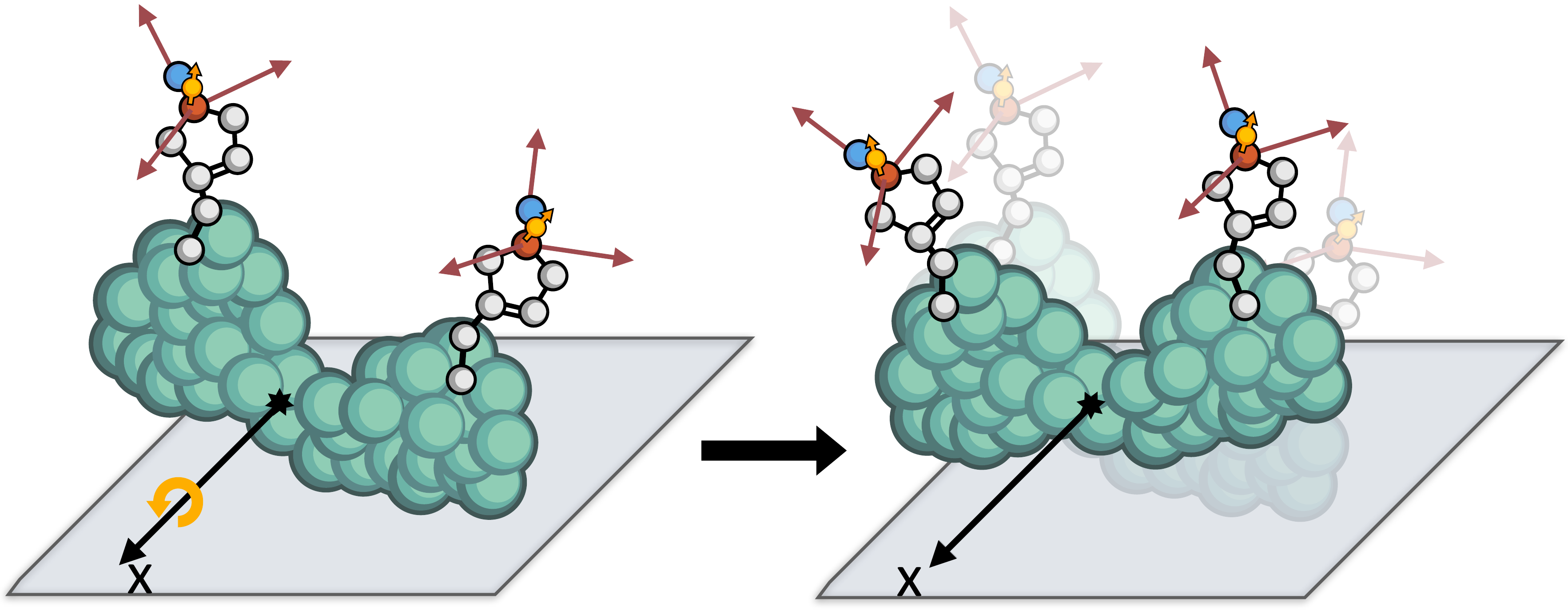}
\caption{Starting from an initial configuration (left), the molecule is slightly rotated around an axis parallel to the laboratory $x$ axis (right). In the right panel, the transparent image shows the initial configuration for better comparison. \label{fig:STumbling}}
\end{figure}

Figure~\ref{fig:STumbling} depicts the effect of molecular tumbling on the orientations and positions of the nitroxides before rotation and after rotation. For all simulations, the molecule rotates around an axis that is parallel to the laboratory $x$ axis and that contains the point $\vec{r}_0 = (3,0,6)$\ nm. For the simulations of Fig.~3(a,d), the equilibrium positions of the two nitroxides are $\vec{r}_{1,{\rm eq}} = (-2.10,2.17,6.24)$\ nm and $\vec{r}_{2,{\rm eq}} = (0.4,0.3,7.3)$\ nm and their equilibrium orientations are $(\theta_{1,{\rm eq}},\varphi_{1,{\rm eq}}) = (11.46,-91.67)^\circ$ and $(\theta_{2,{\rm eq}},\varphi_{2,{\rm eq}}) = (91.67,154.70)^\circ$. For the simulation of Fig.~3(b), the equilibrium positions are $\vec{r}_{1,{\rm eq}} = (-2.10,2.17,6.24)$\ nm and $\vec{r}_{2,{\rm eq}} = (0.89,0.39,8.27)$\ nm and the equilibrium orientations are $(\theta_{1,{\rm eq}},\varphi_{1,{\rm eq}}) = (11.46,-91.67)^\circ$ and $(\theta_{2,{\rm eq}},\varphi_{2,{\rm eq}}) = (91.67,154.70)^\circ$. For the simulation of Fig.~3(c), the equilibrium positions are $\vec{r}_{1,{\rm eq}} = (-2.10,2.17,6.24)$\ nm and $\vec{r}_{2,{\rm eq}} = (0.4,0.3,7.3)$\ nm and the equilibrium orientations are $(\theta_{1,{\rm eq}},\varphi_{1,{\rm eq}}) = (63.03,-91.67)^\circ$ and $(\theta_{2,{\rm eq}},\varphi_{2,{\rm eq}}) = (51.57,154.70)^\circ$. All positions are measured with respect to an origin located at the NV center. The average spectrum $\overline{\langle\sigma^x\rangle}_{\rm NV}$ is obtained by averaging spectra assuming that the rotation angle $\delta$ follows a Gaussian distribution with zero mean and standard deviation $6.25^\circ$. Finally, note that the molecular tumbling has a correlation time of $\sim$1 ms for typical proteins at room temperature~\cite{Shi15}. Since the NV signal must be averaged over several seconds (see Sec.~\ref{sec:bayesInference}), all tumbling configurations are averaged over in the course of the experiment

\subsection{Inference model \label{sec:inferenceModel}}
In this section, we motivate the expression used for Bayesian parameter inference in the main text. We make the following simplifying assumptions:
\begin{enumerate}
\item{The system is simplified to a single NV interacting with a single target nitroxide ($^{14}$N) with orientation $(\theta,\varphi)$. The only role of the other nitroxide is to introduce an energy shift $\pm g_{12}(\beta)/2$ on the target nitroxide depending on its state. Here, $g_{12}(\beta)\propto d_{12}^{-3}(1-3\cos^2\beta)$ is the dipole-dipole coupling and $\beta$ is the angle between the magnetic field and the vector $\vec{r}_{12}$ joining the nitroxides.}
\item{The longitudinal coupling $a^z$ between the NV and target nitroxide stays approximately constant during molecular tumbling. This is a reasonable approximation for small molecular tumbling angles.}
\item{The driving frequency $\omega_{\rm RF}$ is close to resonance with the $E_0(\theta)$ energy-transition branch of the target nitroxide and far off resonance with the other branches. Therefore, non-trivial evolution of the target nitroxide occurs only for nuclear eigenstate $|\widetilde{0}\rangle$.}
\item{The $\pi$-pulse on the NV is simply described by the action of the operator $-i \sigma^x$, i.e., the interaction between the NV and label electrons during irradiation can be neglected.}
\end{enumerate}
With these assumptions, the Hamiltonian of the target nitroxide during free evolution is $H_{{\rm free},\pm} = \left\{\left[E_0(\theta) - \omega_{\rm RF} \pm \frac{g_{12}(\beta)}{2}\right] J^z + \frac{a^z}{2} \sigma^z J^z\right\}\otimes|\widetilde{0}\rangle\langle\widetilde{0}|$ . Similarly, the Hamiltonian of the target nitroxide during irradiation is $H_{{\rm pulse},\pm} = \left\{\left[E_0(\theta) - \omega_{\rm RF} \pm \frac{g_{12}(\beta)}{2}\right]J^z + \Omega_{\rm RF} J^x\right\}\otimes|\widetilde{0}\rangle\langle\widetilde{0}|$. In these expressions, the index $\pm$ labels the states of the other nitroxide electron. For coherent dynamics, these Hamiltonians yield the propagator $U_\pm = e^{-i H_{{\rm free},\pm} \tau_{\rm free}} e^{-i H_{{\rm pulse},\pm} \frac{\pi}{\Omega_{\rm RF}}} (-i \sigma^x) e^{-i H_{{\rm free},\pm} \tau_{\rm free}}$. We assume an initial state $\rho = | + \rangle\langle +| \otimes \frac{\mathbb{I}}{2} \otimes \frac{\mathbb{I}}{3}$, i.e., the target nitroxide is in a fully mixed state. We also assume that the other nitroxide is in a fully mixed state so that the evolutions $U_\pm$ have equal probability. With these initial conditions, we find that the spectrum $\mathcal{S}(\omega_{\rm RF},\theta,\beta) = \langle\sigma^x\rangle_{\rm NV}$ has the form
\begin{align}
\begin{aligned}
\mathcal{S}(\omega_{\rm RF}, \theta, \beta) &\approx \sum_{s=+,-} \frac{1}{2}{\rm Tr}\left(U_s \rho U^\dag_s \sigma^x\right) = \mathcal{S}_0 - \sum_{s=+,-} \mathcal{C}_s \left[\frac{\Omega_{\rm RF}}{\Omega_s(\omega_{\rm RF},\theta,\beta)}\right]^2\sin^2\left[\frac{\pi}{2} \frac{\Omega_s(\omega_{\rm RF},\theta,\beta)}{\Omega_{\rm RF}}\right], \label{eq:ScoherentSpectrum}
\end{aligned}
\end{align}
where $\Omega_\pm^2(\omega_{\rm RF},\theta,\beta) = \Omega_{\rm RF}^2 + \left[\omega_{\rm RF}-E_0(\theta) \pm g_{12}(\beta)/2 \right]^2$, $\mathcal{S}_0 = 1$, and $\mathcal{C}_s = \frac{1}{6}\sin^2\left(\frac{a^z \tau_{\rm free}}{2}\right)$. For a fixed value of $\tau_{\rm free}$, we expect the main effect of dissipation to be two-fold. First, NV decoherence can modify the baseline NV response $\mathcal{S}_0$ when the drive is off-resonant. Second, NV decoherence and relaxation of the nitroxide electrons can reduce the contrast $\mathcal{C}_s$ in a way that may depend on $s = \pm$. Therefore, we leave $\mathcal{S}_0$, $\mathcal{C}_+$, and $\mathcal{C}_-$ as adjustable parameters for the dissipative model. In the presence of molecular tumbling, the average spectrum $\overline{\mathcal{S}}(\omega_{\rm RF})$ is obtained by averaging Eq.~\eqref{eq:ScoherentSpectrum} over a suitable distribution of the tumbling angle $\delta$, with the angles $\theta$ and $\beta$ both depending on $\delta$. For tumbling by an angle $\delta$ around an axis parallel to the laboratory $x$ axis (see Fig.~\ref{fig:STumbling}), the azimuth angle $\theta$ of the target nitroxide is related to $\delta$ by
\begin{align}
\begin{aligned}
\theta(\delta) = \arccos\left[\cos(\delta)\cos(\theta_{\rm eq})+\sin(\delta)\sin(\theta_{\rm eq})\sin(\varphi_{\rm eq})\right],
\end{aligned}
\end{align}
where $(\theta_{\rm eq},\varphi_{\rm eq})$ is the equilibrium orientation of the target nitroxide. Moreover, the dependence of the angle $\beta$ on the tumbling angle $\delta$ is described by the expression $\cos^2[\beta(\delta)]=A_\beta^2\cos^2(\delta + \phi_\beta)$, where $A_\beta$ and $\phi_\beta$ are parameters capturing the orientation of $\vec{r}_{12}$ in the $x-y$ plane. Using these relations, the average spectrum $\overline{\mathcal{S}}(\omega_{\rm RF})$ is obtained by numerically averaging $\mathcal{S}[\omega_{\rm RF},\theta(\delta),\beta(\delta)]$ over $\delta$ assuming that $\delta$ is Gaussian-distributed with unknown standard deviation $\sigma_\delta$. When trying to infer $d_{12}$, there are nine adjustable parameters in total. We assume that $\mathcal{S}_0$ is calibrated and known. The remaining eight adjustable parameters are ${\bf V}=\{\mathcal{C}_+,\mathcal{C}_-,\theta_{\rm eq},\varphi_{\rm eq},A_\beta,\phi_\beta,d_{12},\sigma_\delta\}$.

\subsection{Inference of the distance between nitroxide electron-spin labels}\label{sec:bayesInference}
\begin{figure}[h]
\hspace{-0. cm}\includegraphics[width=0.8\columnwidth]{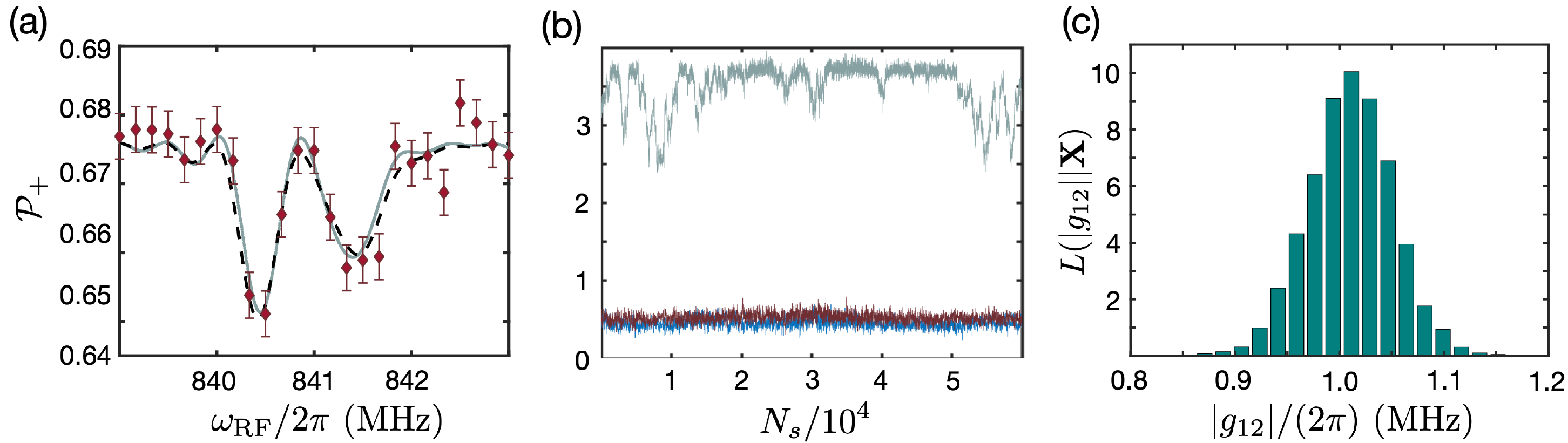}
\caption{(a) Simulated experimental dataset ${\bf X}$ (red points). The dataset contains $M=25$ frequencies $\omega_{{\rm RF},j}$ swept across the target nitroxide resonance, with $N_m = 2\times 10^4$ ideal measurements for each frequency. The error bars are given by $\sigma_m = \sqrt{\left(1-\overline{S}^2\right)/4N_m} \approx \sqrt{0.22/N_m}$. The ``true'' spectrum $\left(\overline{\langle\sigma^x\rangle}_{\rm NV}+1\right)/2$ (solid line) and the simplified model $\left(\overline{\mathcal{S}}(\omega_{\rm RF})+1\right)/2$ (dashed line) are also shown (see text for details about the used parameters). (b) Metropolis Markov chain for the parameters $\mathcal{C}_\pm$ (red and blue) and $d_{12}$ (grey). The contrasts $\mathcal{C}_\pm$ are dimensionless and are multiplied by $10$ for easier viewing. The distance $d_{12}$ is measured in nm. (c) Marginal posterior $L(|g_{12}||{\bf X})$ obtained from the Markov chain by combining the parameters $d_{12}$, $A_\beta$, and $\phi_\beta$ according to $g_{12}=\left(\mu_0 \gamma_e^2 \hbar/4\pi d_{12}^3\right) \left(1-3A_\beta^2\cos^2 \phi_\beta \right)$. All samples before $N_s = 10^4$ were discarded. The expectation value is $|g_{12}|/2\pi=1.010(41)$\ MHz, in good agreement with the ideal value $|g_{12}|/2\pi = 1$\ MHz. The error is given by the standard deviation of the marginal posterior. \label{fig:SInference}}
\end{figure}

We now describe how to infer system parameters from the simplified model discussed in Sec.~\ref{sec:inferenceModel}. We first simulate the acquisition of experimental data from a spectrum obtained numerically using Eq.~\eqref{eq:Slindblad}. Here, we use the spectrum of Fig.~3(a) in the main text. The data has the form ${\bf X}=\{X_1,\ldots,X_M\}$, where $X_j$ is an estimate of the probability $\mathcal{P}_+ = \left(1+\overline{\langle \sigma^x\rangle}_{\rm NV}\right)/2$ of the NV occupying the $+1$ eigenstate of $\sigma^x$ when the target nitroxide is driven at frequency $\omega_{{\rm RF},j}$. Assuming ideal NV measurements, the simulated values of $X_j$ are drawn independently from a Binomial distribution, $X_j \sim B(N_m,\mathcal{P}_+)/N_m$. Fig.~\ref{fig:SInference}(a) shows an example simulated dataset ${\bf X}$. The ``true'' spectrum $\mathcal{P}_+ = \left(1+\overline{\langle \sigma^x\rangle}_{\rm NV}\right)/2$ and the approximate spectrum $\mathcal{P}_+ = \left(1+\overline{\mathcal{S}}(\omega_{\rm RF})\right)/2$ are also shown for comparison. The values of $\theta_{\rm eq},\varphi_{\rm eq}, A_\beta, \phi_\beta, d_{12}$ and $\sigma_\delta$ used to plot the approximate spectrum are the same as for the ``true'' spectrum, while the contrasts $C_{\pm}$ and baseline $\mathcal{S}_0$ are adjusted to fit the ``true'' spectrum. The two spectra are in good agreement, suggesting that our simplified model can be safely used to infer the system parameters ${\bf V}$. To infer the parameters, we assume that $X_j$ is Gaussian distributed with mean $\left(1+\overline{\mathcal{S}}(\omega_{{\rm RF},j})\right)/2$ and variance $\sigma_m^2 = \left(1-\overline{\mathcal{S}}(\omega_{{\rm RF},j})^2\right)/4N_m$. The Gaussian approximation is justifiable when the number $N_m$ of measurements per frequency is large. Since $\overline{\mathcal{S}}\approx 0.34$ at all frequencies for the spectrum of Fig.~3(a), we fix $\sigma_m^2$ to a constant value $\sigma_m^2 \approx 0.22/N_m$. In addition, we assume a uniform prior for ${\bf V}$. The posterior distribution $L({\bf V}|{\bf X})$ is then sampled using a standard Metropolis algorithm~\cite{Gilks96}. Fig.~\ref{fig:SInference}(b) shows an example Markov chain for the parameters $d_{12}$, $\mathcal{C}_+$, and $\mathcal{C}_-$ (the contrasts $\mathcal{C}_\pm$ are multiplied by a factor $10$ for easier viewing). The resulting marginal posterior for $d_{12}$ is shown in Fig.~4(d) of the main text. For reference, the marginal posterior of $|g_{12}|$ is also shown in Fig.~\ref{fig:SInference}(c). The posteriors were obtained by discarding the first $10^4$ steps as a burn-in phase. The Metropolis acceptance rate was $\sim 30\%$.

When the measurements are not ideal, preparation of the NV in the $m_S = 0$ state (mapped to $\sigma^x = +1$) leads to a photon detection with probability $p \ll 1$, while preparation of the NV in the $m_S = \pm 1$ state (mapped to $\sigma^x = -1$) leads to no photon detection. Under these assumptions, we estimate that the measurement noise variance scales as $\sigma_m^2 \approx (1+\overline{\mathcal{S}})/2pN_m$~\cite{Danjou14}. For $\overline{\mathcal{S}}\approx 0.34$ and for an experimentally achievable ground state detection efficiency $p \approx 0.12$~\cite{Wan18}, this gives $\sigma_m^2 \approx 5.6/N_m$. Comparing with the ideal case, we find that the same variance as for ideal measurements is achieved with approximately $25$ times more measurements. In the present case, this would require $N_m \approx 5\times 10^5$ non-ideal measurements per RF frequency. The execution time of the sequence ($4.6\ \mu$s), plus measurement and reinitialization ($\sim3\ \mu$s) and some additional time to assure reliable thermalization of the labels ($\sim3 T_1$) leads to a total time of $\sim20\ \mu$s per shot. Assuming $M = 25$ inspected frequencies and $N_m \approx 5\times 10^5$, the experiment would require 250 s to be completed.

\end{document}